\documentclass{jetpl}
\twocolumn

\lat
\usepackage[dvips]{graphicx}
\usepackage{epsfig,color}

\title{Cluster magnetism of Ba$_4$NbMn$_3$O$_{12}$: localized electrons or molecular orbitals?}

\rtitle{Magnetism of Ba$_4$NbMn$_3$O$_{12}$: localized electrons or molecular orbitals?}

\sodtitle{Magnetism of Ba$_4$NbMn$_3$O$_{12}$: localized electrons or molecular orbitals?}

\author{S.\,V.\,Streltsov$^{a,b,}$\/\thanks{e-mail: streltsov@imp.uran.ru},
D.\,I.\,Khomskii$^{*c}$}

\rauthor{S.\,V.\,Streltsov, D.\,I.\,Khomskii}

\sodauthor{Streltsov, Khomskii}

\address{$^a$Ural Federal University, Mira Str. 19, 620002 Ekaterinburg, Russia\\
~\\$^b$Institute of Metal Physics, Russian Academy of Science, S. Kovalevskaya Str. 18, 620041 Ekaterinburg, Russia\\
~\\$^c$II. Physikalisches Institut, Universit$\ddot a$t zu K$\ddot o$ln,
Z$\ddot u$lpicher Stra$\ss$e 77, D-50937 K$\ddot o$ln, Germany}

\dates{\today}{*}

\abstract{
Recently synthesized Ba$_4$NbMn$_3$O$_{12}$ belong to cluster magnets - systems with tightly bound groups of magnetic ions, in this case Mn$_3$ trimers.  Often such magnetic clusters can be described by molecular orbitals (MO), however strong electron correlations may invalidate this description. To understand the electronic and magnetic state of  Ba$_4$NbMn$_3$O$_{12}$ we carried out ab initio calculations and show that  this system is better described not in MO picture, but as a system with electrons localized on the Mn ions, with strong intra-cluster and weaker inter-cluster exchange. The calculated spin of the Mn$_3$ trimer is $S=2$, in agreement with the experiment.  The predicted magnetic structure of Ba$_4$NbMn$_3$O$_{12}$ is that of ferromagnetic layers of Mn$_3$ trimers, stacked antiferromagnetically.
}


\begin{document}

\maketitle

Typical strong magnets are transition metal (TM) or rare earth (RE) compounds, and the ``carriers'' of magnetic moments are TM or RE ions with strongly correlated electrons\cite{Khomskii-book}. There exist however many materials in which such ions form relatively tightly bound clusters - dimers, trimers, or bigger objects. Electron hopping within such clusters  $t$  can be sufficiently large - comparable or even larger than the respective intra-atomic interactions: Hubbard repulsion $U$ and Hund's coupling $J_H$. In such cases the electronic state of such clusters may be described by molecular orbitals (MO), and electrons would occupy such MO levels. Depending on the electron filling, such clusters can still  have localized magnetic moments, and the solid composed of such clusters may develop magnetic ordering. But in this case the ``carriers'' of magnetism would be not isolated TM ions, but rather such clusters. We can speak in this case of ``molecules in solids'',  cluster magnetism, or cluster Mott insulators\cite{Streltsov-UFN}.  These systems have everything the usual Mott insulators have (electrons localized on such clusters may be localised due to the ``on-cluster'' electron repulsion, substituting the Hubbard $U$ in the usual Hubbard model; they can hop from cluster to cluster and become itinerant, etc). But, besides all the features the usual correlated electron system have, there are extra intracluster degrees of freedom – which can lead for example to charge ordering (e.g. electrons occupying always the ``top'' site of a triangle), they can also have extra orbital ordering (due to possible degeneracy of some molecular orbitals), and other specific features. The study of such ``cluster'' systems present definite interest and is starting to attract more and more attention. 

A few examples of such systems are for instance: materials like Ba$_3$M'M$_2$O$_9$\cite{Ziat2017,Nag2016}, with metals M (Ru or Ir) forming dimers of face-sharing MO$_6$ octahedra. There may be also edge-sharing dimers, e.g., in Li$_2$RuO$_3$\cite{Miura2007,Park2016} or  Y$_5$Mo$_2$O$_{12}$\cite{Torardi1985}. Some of the ``best'' spin liquid materials - molecular crystals\cite{Zhou2017}  also have spins 1/2 localized on dimers. Many materials having Peierls or spin-Peierls transition may be treated as systems of this type. 

Such clusters may be trimers: linear trimers e.g. in Ba$_4$Ru$_3$O$_{10}$\cite{Streltsov2012a} or Ba$_5$AlIr$_2$O$_{11}$\cite{Terzic2015}, or ``flat'' triangular clusters, e.g. Mo$_3$ clusters in Zn$_2$Mo$_3$O$_8$ or Fe$_2$Mo$_3$O$_8$\cite{Kurumaji2017} (which may be doped, e.g., LiZn$_2$Mo$_3$O$_8$\cite{Sheckelton2012}); or V$_3$ clusters formed in LiVO$_2$ below structural transition at $\sim$540 K\cite{Katayama2009}.

Clusters of four TM ions are met e.g. in CaV$_4$O$_9$ (planar plaquettes \cite{Starykh1996}) or in lacunar spinels (tetrahedral clusters in GaV$_4$O$_8$; GaTa$_4$O$_8$) which attracted a lot of attention because of their multiferroic properties (GaV$_4$S$_8$\cite{Nikolaev2018})  or superconductivity  (GaTa$_4$S$_8$\cite{Abd-Elmeguid2004}). Also the famous COSO (Cu$_2$OSeO$_3$) - the first insulating material possessing skyrmions - also contains clusters of four Cu ions, and each such cluster is a carrier of  a spin   1\cite{Janson2014}.

Even bigger clusters were proposed in the literature, e.g., heptamers (clusters of 7 V ions) in AlV$_2$O$_4$\cite{Matsuno2003} (although more recent data show that here we are dealing rather with trimers (triangles)  and tetramers (tetrahedra) of V\cite{Browne2017}, see also Ref.~\cite{Talanov2018}).  Even much larger objects can be considered in this picture - for example the famous buckyballs C$_{60}$\cite{Schon2000}. And of course  also typical molecular magnets, showing quantum tunneling of magnetization,  belong to this category.
\begin{figure}[t!]
 \centering
 \includegraphics[clip=false,width=0.45\textwidth]{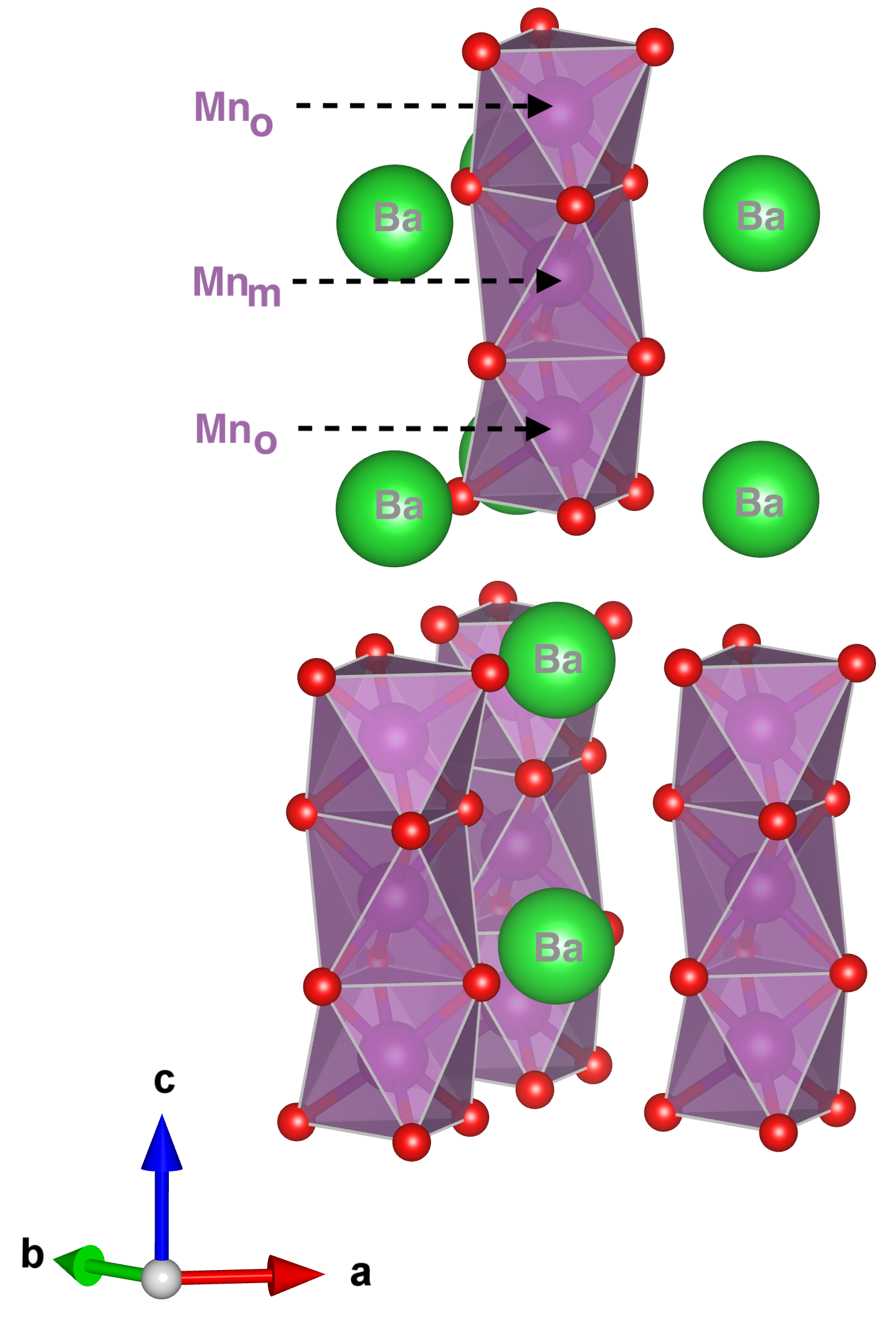}
\caption{\label{crystal-structure}Fig. 1. Crystal structure of Ba$_4$NbMn$_3$O$_{12}$. The Mn ions (violet balls) are in the oxygen (small red balls) octahedra. Three nearest MnO$_6$ octahedra form a trimer sharing their faces. The Mn ions, which are in the middle of the trimer, are labeled as Mn$_m$ throughout the text, while Mn$_o$ are outer Mn ions in the trimer. Ba (large green balls) ions sit in the voids and Nb ions are not shown.}
\end{figure}

In dealing with specific material containing such structural clusters, the first question arising is whether we should consider such materials as composed of TM ions as the main units, or the better description is that of MOs on a cluster. The situation may be different from system to system, not only depending on the specific geometry of the material, but also it can differ for different TM metals. Thus, $3d$ electrons are usually more localized, have smaller intersite hopping, but larger  Hubbard repulsion than, for example, $4d$ and $5d$, thus they may behave as more localized, and one might expect better conditions for the formation of MO states in $4d$ and $5d$ systems than in $3d$ ones  (although of course such MO states are not excluded also in some $3d$ systems). In this respect it is interesting to compare materials with the same crystal structure, but with different TM ions, $3d$ vs. $4d$ or $5d$. An interesting example of such systems are materials with the general formula Ba$_4$M'M$_3$O$_{12}$, with different metals M' occupying single octahedra, and, more important, with the possibility to put either $3d$ ions, e.g. Mn,  or $4d$ (Ru) and $5d$ elements (Ir) at M sites forming face-sharing trimers, see Fig.~\ref{crystal-structure}.  Recently two such materials were synthesized by the group of R. Cava \cite{Nguyen2018a,Nguyen2018}. There exist also similar systems with Ir trimers, e.g. BaIrO$_3$\cite{Brooks2005}, Ba$_3$RIr$_3$O$_{12}$\cite{Shimoda2010}, Ba$_5$CuIr$_3$O$_{12}$ \cite{Ye2018}. And whereas for Ru and Ir trimers most probably the MO picture is a good starting point\cite{Streltsov2012a}, for $3d$ system Ba$_4$NbMn$_3$O$_{12}$\cite{Nguyen2018} the situation is not so obvious.

The structure of Ba$_4$NbMn$_3$O$_{12}$  is shown in Fig. \ref{crystal-structure}. The main interesting for us building block is a linear trimer of three face-sharing MnO$_6$ octahedra. There are two crystallographically different Mn ions: those siting in the middle of the trimer (Mn$_m$) and outer Mn ions (Mn$_o$). The average valence of Mn here is Mn$^{3\frac 23 +}$, i.e. it nominally contains two Mn$^{4+}$ ($d^3$) and one Mn$^{3+}$ ($d^4$). In principle they could form charge-ordered (CO) state, but structural data \cite{Nguyen2018} do not show any indication for that - at least not strong CO. The system is magnetic, with the Curie-Weiss susceptibility, and the most interesting result is that the effective moment $\mu_{eff} = 4.89 \mu_B$ corresponds to spin $S_{tot}=2$ per Mn$_3$ trimer~\cite{Nguyen2018}.

There are different possibilities to get such $S=2$ state of a trimer. The first one (1), discussed in \cite{Nguyen2018}, is that in which two Mn$^{3+}$ ions with the configuration $t_{2g}^3$, e.g., at the edges of a trimer, have spins $S_z = +3/2$,  and Mn$^{3+}$ in the center is in the low-spin configuration $t_{2g}^4$ with spin $S_z = -1$, so that the total spin of a trimer is $S_{tot}=2$. But the low-spin state for Mn$^{3+}$ ions is very uncommon.

Another possibility (2) is the state with localized electrons and with the $t_{2g}^3$ occupation for all three sites,  the remaining electron occupying the $e_g$ orbitals is delocalized over the whole trimer. In this case $t_{2g}$ shells of Mn ions may have spins +3/2, -3/2, +3/2, and due to intra-atomic exchange coupling the $e_g$ electron should have spin corresponding to the total spin of $t_{2g}$ electrons, i.e. it will have spin up, $S=+1/2$. As a result we would again have $S_{tot} =2$, but without assuming the low-spin state of Mn$^{3+}$.
\begin{figure}[t!]
 \centering
 \includegraphics[clip=false,width=0.5\textwidth]{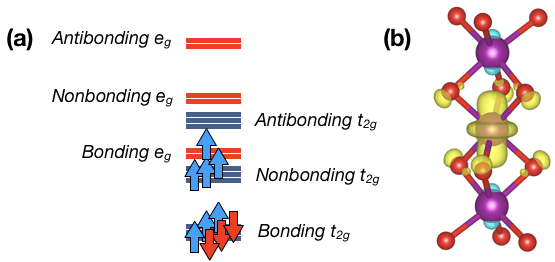}
\caption{\label{level-diagram}Fig. 2.  (a) Possible level diagram of a trimer formed by TM ions. Here we assume that bonding $e_g$ orbitals are higher than nonbonding $t_{2g}$ due to substantial $t_{2g}-e_g$ splitting. (b) The spin-density corresponding to the band centred, at $\sim -7.5$ eV in Fig.~\ref{DOS}.}
\end{figure}

And finally, (3) in principle all electrons, both $t_{2g}$ and $e_g$ ones, may form MO states. Direct overlap of the $t_{2g}$ orbitals in a common face geometry may be significant,  cf. for example\cite{Kugel2015,Khomskii2016}. It is easy to show \cite{Streltsov2012a} that in this case both the $t_{2g}$ and $e_g$ states form bonding, nonbonding, and antibonding orbitals and the level diagram can be sketched as shown in Fig.~\ref{level-diagram}(a). We have to fill these levels by 9 $t_{2g}$ electrons. The Hund's rule would then give the occupations shown in Fig.~\ref{level-diagram}(a), i.e. $t_{2g}$ subsystem would have total spin $S_{t_{2g}} = 3/2$. The remaining $e_g$ electron would also form MO state, with the same spin by Hund's rule, i.e. the total spin of a trimer would again be $S_{tot}=2$.

To check which of these three possibilities is indeed realised in Ba$_4$NbMn$_3$O$_{12}$, we carried our {\it ab initio} calculations using pseudo-potential VASP code\cite{vasp}. Perdew-Burke-Enzerhof variant of the generalized gradient approximation (GGA)\cite{Perdew1996} was chosen. The crystal structure was taken from Ref.~\cite{Nguyen2018}. Integration over the Brillouin zone was performed on the mesh of $7 \times 7 \times 7$  $\mathbf k-$points. To take into account electronic correlations we used GGA+U method as formulated in Ref.~\cite{Liechtenstein1995} with $U=4.5$ eV and $J_H=0.9$ eV \cite{Streltsov2008} (we also checked that variation of $U$ on $\pm 1$ eV does not change the results). 

The lowest in energy magnetic configuration is the one with spins on middle and outer Mn ions ordered antiferromagnetically (i.e. $\uparrow  \downarrow \uparrow$). The FM state ($\uparrow \uparrow\uparrow$) is 116 meV/f.u higher in energy. Thus, we see that there is a strong antiferromagnetic exchange coupling between nearest Mn ions. This coupling is due to both direct and oxygen assisted $t_{2g}-t_{2g}$ hoppings. As we will show below the $e_g$ states are partially filled in Ba$_4$NbMn$_3$O$_{12}$ and one may expect FM double exchange-like, contribution, but our calculations show that the total exchange interaction is AFM and thus direct and superexchange due to half-filled $t_{2g}$ orbitals overwhelms the double exchange.
\begin{figure}[t!]
 \centering
 \includegraphics[clip=false,width=0.45\textwidth]{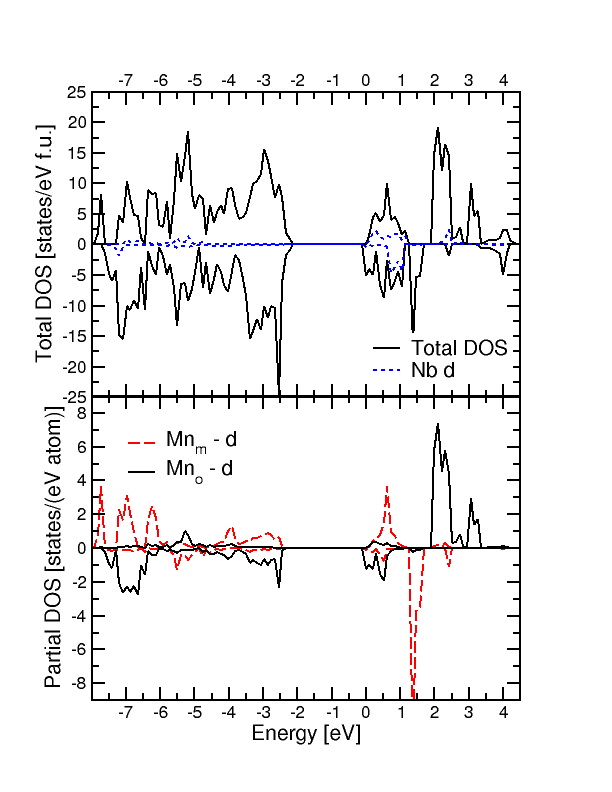}
\caption{\label{DOS}Fig. 3.  Total and partial densities of states (DOS) for Ba$_4$NbMn$_3$O$_{12}$ obtained in the GGA+U calculation. Fermi level is at zero. The system is nearly insulating in our calculations with small DOS at the Fermi level, probably
due to not sufficiently accurate treatment of the empty Nb $4d$ band in the GGA.}
\end{figure}

We proceed with the analysis of the lowest in energy $\uparrow  \downarrow \uparrow$ configuration. Both types of the Mn ions have nearly the same number of $d-$electrons  $\delta n = n_{Mn_m} - n_{Mn_o} = 0.05$. Thus, there is no charge-ordering in Ba$_4$NbMn$_3$O$_{12}$. The total magnetic moment is 3.9 $\mu_B$ per formula unit (f.u.), which is very close to what one might expect for $S_{tot}=2$ per trimer and to experimental results. Further analysis of the occupation matrices shows that all $t_{2g}$ states are half-filled and provide $m^{t_{2g}}_{Mn_m} = -2.5 \mu_B$ and $m^{t_{2g}}_{Mn_o} = 2.6 \mu_B$ to total magnetic moments on these ions, which were found to be $m_{Mn_m} = -2.9 \mu_B$ and $m_{Mn_o} = 3.3 \mu_B$.  The difference between $t_{2g}-$only and total magnetic moments is due to spin polarization of the $e_g$ shell. The spin density corresponding to 1 $\mu_B$ is spread over $e_g$ orbitals (Mn$_m$ has $m^{e_g}_{Mn_m} = -0.4 \mu_B$, each Mn$_o$: $m^{e_g}_{Mn_o} = 0.7 \mu_B$). This shows that there is no low-spin state of Mn$^{3+}$ (which would correspond to the $t_{2g}^4$ state, without any $e_g$ contribution), i.e. the scenario (1) is apparently not realised here.

We get further insight as to the details of electronic structure (which affects magnetic properties) from the density of states and spin density plots shown in Figs.~\ref{DOS} and \ref{level-diagram}(b). The spin density in Fig.~\ref{level-diagram}(b) corresponds to the lowest in energy $d$ band at $-7.5$ eV. We see that this spin density  is mostly due to the $t_{2g}$ orbitals (in fact $a_{1g}$), and it is centred on one of Mn ions and therefore scenario (2) with the site-localized $t_{2g}$ electrons looks plausible. The $e_g$ electrons in this picture are spread over the trimer. 

Thus according to our calculations Ba$_4$NbMn$_3$O$_{12}$ is apparently much closer to the localized limit, in contrast to many $4d$ and $5d$ systems. Indeed, scenario (3) with pure MO state of a trimer, for both $t_{2g}$ and $e_g$ electrons  does not seem to correspond to what we have in the GGA+U calculations. First of all, in this picture one would expect the formation of well-separated bonding and antibonding bands. However, the obtained electronic structure, see Fig.~\ref{DOS}, does not show such bands - clearly seen for example in similar calculations of Ba$_4$Ru$_3$O$_{10}$\cite{Streltsov2012a}. The second, more important and easier to follow argument is that in the MO picture the spin polarization of all $t_{2g}$ electrons would be positive on {\it any} Mn sites (bonding orbitals are completely filled and do not contribute to the total magnetization on any site, whereas spin moment is provided by nonbonding orbitals only, see Fig. 2a). However, we have seen that in fact different Mn ions have different spin orientations in the GGA+U calculation. 

\begin{figure}[t!]
 \centering
 \includegraphics[clip=false,width=0.5\textwidth]{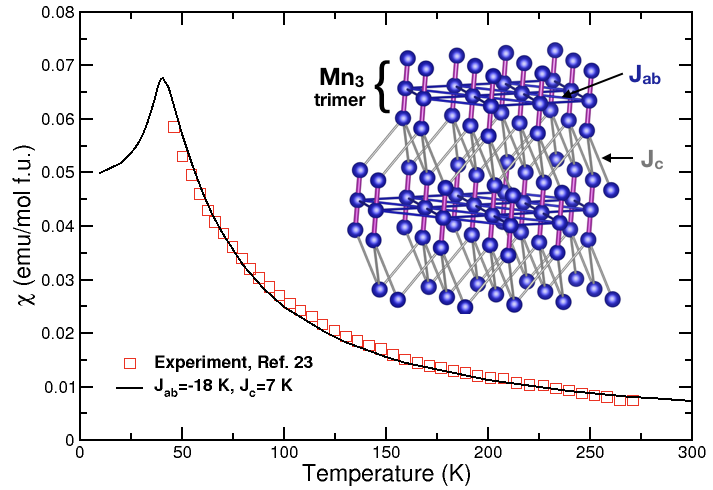}
\caption{\label{chi}Fig. 4.  Temperature dependence of magnetic susceptibility as obtained in the classical Heisenberg model of coupled clusters of $S=2$ (the Mn trimer) on the lattice shown in the inset. For comparison experimental results taken from Ref.~\cite{Nguyen2018} are shown.}
\end{figure}

Finally, in order to test possibility of cluster magnetism in Ba$_4$NbMn$_3$O$_{12}$ we simulated its magnetic susceptibility, $\chi(T)$ using classical Heisenberg model $H = \sum_{i>j} J_{ij} \vec S_i \vec S_j$. The spin model consists of $S=2$ trimers forming triangular planes with in-plane exchange $J_{ab}$, coupled by the diagonal interlayer  interaction $J_c$ as shown in the inset of Fig.~\ref{chi}. We used classical Monte-Carlo (MC) method employing local and cluster updates with the box size $10 \times 10 \times 10$ as realized in the ALPS library\cite{ALPS} with interface generated by the JaSS code\cite{Jass}. Number of sweeps in the MC run was chosen to be 10 000 000. Because of too large unit cell we have not calculated intercluster exchanges directly, but fitted them to experimental magnetic susceptibility (corrected by temperature independent contribution $\chi_0$)\cite{Nguyen2018} with chosen spin model. The best fitting was achieved for $J_{ab} = -18$ K (FM) and $J_{c}=7$ K (AFM). Thus, we see that the model of exchange-coupled $S=2$ trimers may describe observed temperature dependence of magnetic susceptibility. From these results we can deduce that the long-range magnetic ordering should be such that with the triangular layer Mn$_3$ trimers in Ba$_4$NbMn$_3$O$_{12}$ is ordered ferromagnetically, the neighbouring such layers being antiparallel (similar to A-type antiferromagnetic ordering in perovskites).  

Summarizing, we showed that Ba$_4$NbMn$_3$O$_{12}$ can be considered as a cluster magnet. However, in contrast to many similar systems with the $4d$ and $5d$ electrons, for which the MO description applies\cite{Streltsov2012a,Ye2018}, it is not the case for the considered $3d$ material Ba$_4$NbMn$_3$O$_{12}$. This material should rather be described by the picture of electrons localised on TM sites (here Mn), which, being strongly antiferromagnetically coupled, form $S=2$ Mn$_3$ trimers. Apparently the inter-cite hopping here is not large enough to overcome the strong on-site Hubbard and Hund interactions.  Of course these two pictures, of completely localized electrons or  pure MO state of Fig.~\ref{level-diagram}(a), are the limiting cases. The real situation is always somewhere in between (which actually helps to form total moment $S_{tot}=2$). We also predict that the ordered magnetic structure of this system would be that of ferromagnetic layers of Mn$_3$ trimers stacked antiferromagnetically.

We are grateful to R. Cava for informing us about results of \cite{Nguyen2018} prior to publication and for stimulating discussions. Calculations of electronic and magnetic properties Ba$_4$NbMn$_3$O$_{12}$ were supported by the Russian Scientific Foundation (project no. 17-12-01207). D.I. Khomskii acknowledges support of CRC 1238.


\begin{thebibliography}{99}

\bibitem{Khomskii-book} D. I. Khomskii, Transition Metal Compounds (Cambridge University Press, 2014).

\bibitem{Streltsov-UFN} S. V. Streltsov and D. I. Khomskii, Physics-Uspekhi {\bf 60}, 1121 (2017).

\bibitem{Ziat2017} D. Ziat, A. A. Aczel, R. Sinclair, Q. Chen, H. D. Zhou, T. J. Williams, M. B. Stone, A. Verrier, and J. A. Quilliam, Phys. Rev. B {\bf 95}, 184424 (2017)

\bibitem{Nag2016} A. Nag, S. Middey, S. Bhowal, S. K. Panda, R. Mathieu, J. C. Orain, F. Bert, P. Mendels, P. G. Freeman, M. Mansson, et al., Phys. Rev. Lett. {\bf 116}, 097205 (2016)

\bibitem{Miura2007} Y. Miura, Y. Yasui, M. Sato, N. Igawa, and K. Kaku- rai, Journal of the Physical Society of Japan {\bf 76}, 033705 (2007)

\bibitem{Park2016} J. Park, T.-y. Tan, D. T. Adroja, A. Daoud-Aladine, S. Choi, D.-y. Cho, S.-h. Lee, J. Kim, H. Sim, T. Morioka, et al., Scientific reports {\bf 6}, 25238 (2016)

\bibitem{Torardi1985}C. C. Torardi, C. Fecketter, W. H. McCarroll, and F. J. DiSalvo, Journal of Solid State Chemistry {\bf 60}, 332 (1985).

\bibitem{Zhou2017} Y. Zhou, K. Kanoda, and T.-k. Ng, Rev. Mod. Phys. {\bf 89},
025003 (2017).

\bibitem{Streltsov2012a} S. V. Streltsov and D. I. Khomskii, Phys. Rev. B
{\bf 86}, 064429 (2012)

\bibitem{Terzic2015} J. Terzic, J. C. Wang, F. Ye, W. H. Song, S. J. Yuan, S. Aswartham, L. E. DeLong, S. V. Streltsov, D. I. Khomskii, and G. Cao, Phys. Rev. B {\bf 91}, 235147 (2015).

\bibitem{Kurumaji2017} T. Kurumaji, Y. Takahashi, J. Fujioka, R. Masuda, H. Shishikura, S. Ishiwata, and Y. Tokura, Phys. Rev. Lett. {\bf 119}, 077206 (2017)

\bibitem{Sheckelton2012} J. P. Sheckelton, J. R. Neilson, D. G. Soltan, and T. M. McQueen, Nature Materials {\bf 11}, 493 (2012).

\bibitem{Katayama2009} N. Katayama, M. Uchida, D. Hashizume, S. Niitaka, J. Matsuno, D. Matsumura, Y. Nishihata, J. Mizuki, N. Takeshita, A. Gauzzi, et al., Phys. Rev. Lett. {\bf 103}, 146405 (2009)

\bibitem{Starykh1996} O. A. Starykh, M. E. Zhitomirsky, D. I. Khomskii, R. R. P. Singh, and K. Ueda, Phys. Rev. Lett. {\bf 77}, 2558 (1996) 

\bibitem{Nikolaev2018} I. Kezsmarki, S. Bordacs, P. Milde, E. Neuber, L. M. Eng, J. S. White, H. M. Ronnow, C. D. Dewhurst, M. Mochizuki, K. Yanai, H. Nakamura, D. Ehlers, V. Tsurkan, and A. Loidl, Nature Materials {\bf 14}, 1116 (2015).

\bibitem{Abd-Elmeguid2004} M. M. Abd-Elmeguid, B. Ni, D. I. Khomskii, R. Pocha, D. Johrendt, X. Wang, and K. Syassen, Phys. Rev. Lett. {\bf 93}, 126403 (2004) 

\bibitem{Janson2014} O. Janson, I. Rousochatzakis, A.A. Tsirlin, M. Belesi, A.A Leonov, U.K. RŁoler, J. Van den Brink, and H. Rosner, Nature Communications {\bf 5}, 5376 (2014). 

\bibitem{Matsuno2003} K. Matsuno, T. Katsufuji, S. Mori, M. Nohara, A. Machida, Y. Moritomo, K. Kato, E. Nishibori, M. Takata, M. Sakata, et al., Phys. Rev. Lett. {\bf 90}, 7 (2003)

\bibitem{Talanov2018} M. V. Talanov, V. B. Shirokov, L. A. Avakyan, V. M. Talanov, and K. S. Borlakov, Acta Crystallogr. B {\bf 74}, 1 (2018).

\bibitem{Browne2017} A. J. Browne, S. A. J. Kimber, and J. P. Attfield, Phys. Rev. Mat. {\bf 1}, 052003 (2017).

\bibitem{Schon2000} H. W. Kroto, J. R. Heath, S. C. O'Brien, R. F. Curl and R. E. Smalley, Nature {\bf 318}, 162 (1985).

\bibitem{Nguyen2018a} L. T. Nguyen, T. Halloran, W. Xie, T. Kong, C. L. Broholm, and R. J. Cava, Physical Review Materials {\bf 2}, 054414 (2018).

\bibitem{Nguyen2018} L. Nguyen, T. Kong, and R. Cava, arXiv:1810.00763

\bibitem{Brooks2005} M. L. Brooks, S. J. Blundell, T. Lancaster, W. Hayes,
F. L. Pratt, P. P. C. Frampton, and P. D. Battle, Phys.
Rev. B {\bf 71}, 220441 (2005).

\bibitem{Shimoda2010} Y. Shimoda, Y. Doi, M. Wakeshima, and Y. Hinatsu,
Journal of Solid State Chemistry {\bf 183}, 1962 (2010)

\bibitem{Ye2018} M. Ye, H.-S. Kim, J.-W. Kim, C.-J. Won, K. Haule, D. Vanderbilt, S.-W. Cheong, and G. Blumberg, ArXiv e-prints (2018), 1808.10407.

\bibitem{Kugel2015} K. I. Kugel, D. I. Khomskii, A. O. Sboychakov, and S. V. Streltsov, Phys. Rev. B {\bf 91}, 155125 (2015)

\bibitem{Khomskii2016} D. I. Khomskii, K. I. Kugel, A. O. Sboychakov, and S. V. Streltsov, Journal of Experimental and Theoretical Physics {\bf 122}, 484 (2016).

\bibitem{vasp} G. Kresse and J. Furthm\"uller, Phys. Rev. B {\bf 54}, 11169 (1996).

\bibitem{Perdew1996} J. P. Perdew, K. Burke, and M. Ernzerhof, Phys. Rev. Lett. {\bf 77}, 3865 (1996)

\bibitem{Liechtenstein1995}  A. I. Liechtenstein, V. I. Anisimov, and J. Zaanen, Phys. Rev. Lett. {\bf 52}, 5467 (1995)

\bibitem{Streltsov2008} S. V. Streltsov and D. I. Khomskii, Phys. Rev. B {\bf 89}, 201115 (2014).

\bibitem{ALPS} B. Bauer et al., J. Stat. Mech. Theory Exp. 05001 (2011).

\bibitem{JaSS} S. Streltsov et al., www.jass-code.org




\end{thebibliography}
\end{document}